\documentclass{aastex}
\usepackage{emulateapj5,psfig}

\newcommand{\tm}{12$\mu$m}
\newcommand{\firr}{$f_{25}/f_{60}$}
\newcommand{\ftv}{$f_{25}$}
\newcommand{\fst}{$f_{60}$}
\newcommand{\frir}{$S_{20cm}/f_{60}$}
\newcommand{\hxir}{$HX/f_{60}$}
\newcommand{\ha}{H$\alpha$}
\newcommand{\hb}{H$\beta$}

\slugcomment{Submitted 2001 March 20; accepted 2001 May 08}
\journalinfo{Accepted for publication in ApJ Letters}

\shortauthors{Tran}
\shorttitle{Obscured Seyfert 2 Galaxies}

\begin{document}

\title{Hidden Broad Line Seyfert 2 Galaxies in the C\lowercase{f}A and \lowercase{12$\mu$m} Samples}

\author{Hien D. Tran}
\affil{Department of Physics and Astronomy, Johns Hopkins University, Baltimore, MD 21218}
\email{tran@pha.jhu.edu}

\begin{abstract}
We report the results of a spectropolarimetric survey of the
CfA and 12$\mu$m samples of Seyfert 2 galaxies (S2s). Polarized (hidden) broad line regions
(HBLRs) are confirmed in a number of galaxies, and several new cases (F02581--1136, 
MCG -3-58-7, NGC 5995, NGC 6552, NGC 7682) are reported. 
The \tm~S2 sample shows a significantly higher incidence of HBLR (50\%) than its CfA counterpart (30\%),
suggesting that the latter may be incomplete in hidden AGNs. Compared to the non-HBLR S2s,
the HBLR S2s display distinctly higher radio power relative to their far-infrared output
and hotter dust temperature as indicated by the \firr~color. However,  
the level of obscuration is indistinguishable between the two types of S2. 
These results strongly support the existence of two intrinsically different populations of 
S2: one harboring an energetic, hidden S1 nucleus with BLR, and the other, a ``pure S2'', with weak or
absent S1 nucleus and a strong, perhaps dominating starburst component. 
%There is also evidence that the fraction of HBLR increases with radio loudness of the AGN.
Thus, the simple purely orientation-based unification model is not
applicable to {\it all}~Seyfert galaxies. 
%and evolutionary processes may likely be at work.
\end{abstract}


\section{Introduction} \label{intro}

It is sobering to realize that nearly two decades after the ground-breaking observations of NGC 1068 by 
\citet{ma83} showing that some Seyfert 2 galaxies (S2) are basically the same class of object as Seyfert 1s (S1) 
but viewed from a different direction, we still don't know for certain if this ``unified model'' (UM) is 
indeed applicable to {\it all} Seyfert galaxies. 
Besides NGC 1068, there have been plenty of other examples of polarized (hidden) broad line regions seen 
in reflected light (HBLR) in nearly all types of active galactic nuclei (AGNs; see review by Antonucci 2001). 
Nevertheless, while there is no question that this UM applies to {\it some} type 2 AGNs, 
the questions still remain:
Are {\it all} S2s the same as S1s seen from a different line of sight?
If not, what is the fraction of Seyfert galaxies containing S1? 
And what is the exact nature of those S2s that do not show any HBLRs: 
is it an intrinsic property (i.e., ``pure'' S2), or are the HBLRs too obscured to be detected?

Several polarimetric surveys have attempted to address some of these questions 
\citep{mg90,kay94,hlb97,mor00}. Common problems often faced include:
(i) severe biases in the polarized galaxy sample; to maximize the use of telescope time, most early spectropolarimetric
targets were chosen simply because they were known to be polarized, and
(ii) number of objects too small for statistical study; since polarimetric observations are time consuming and tedious, 
it is difficult to find a suitably large sample of S2s with HBLR. 
What is needed is a systematic search of a large, complete and unbiased sample of S2s.
In this Letter, we report the main results and implications of a large survey, of two of the most complete,
and best studied samples of Seyfert galaxies: the CfA \citep{hb92} and the extended \tm~(Rush, Malkan, \& Spinoglio 
1993, RMS93) samples. 
The CfA sample has been revered as the most complete and unbiased sample of nearby Seyferts that are optically
selected (however, see below). 
The extended \tm~sample has the advantage of being significantly larger, 
thereby enabling better statistical analysis, and by selecting in the mid-infrared (mid-IR), it is more suitable for 
detecting buried AGNs that might have been missed by optical selection technique alone. Comparison of the two samples 
should prove most revealing. 

\section{Observations and Results} \label{obs}

We carried out the spectropolarimetric observations of S2s selected from the CfA and
\tm~samples mostly at Lick and Palomar Observatories, with a few
at Keck Observatory, during 13 runs between 1993 December and 2000 January. 
Detailed description of the observations and data will be presented in a separate paper (paper II). 
 Typically, each target was observed through the usual polarimetric sequence
of 4 $\times$~900s or 4 $\times$~1200s, and repeated when necessary to improve S/N. 
For objects observed at Keck, the typical exposure time was 4 $\times$~300s. 
Since the sample objects all have comparable brightness, these observations probe to similar level of sensitivity 
for the entire sample. We also supplement our observations with results from previous surveys. 

Observation of the CfA sample is complete. We observed only those classified as S2 by Osterbrock \& Martel (1993, OM93). 
Intermediate Seyferts (i.e., 1.8, 1.9) were excluded. One object, Mrk 461, for which OM93 did not obtain spectrum, 
turns out to be a H II galaxy based on our observations, and thus will be discarded from the Seyfert consideration. 
In total, 14 S2s were observed from the CfA sample. 

For the \tm~sample, the survey is essentially complete. 
In addition to those in Table 2 of RMS93, we also include one additional object (MCG -4-2-18) from their starburst (SB)
galaxy list (their Table 3) and three objects (NGC 3147, 3822, 6552) from their ``normal'' galaxy list (their Table 5). 
Spectroscopic observations of these sources reveal that they clearly show high ionization spectra worthy 
%TABLE 1 
\hspace*{-0.8cm}
\scalebox{0.96}{\includegraphics[160,58][455,784]{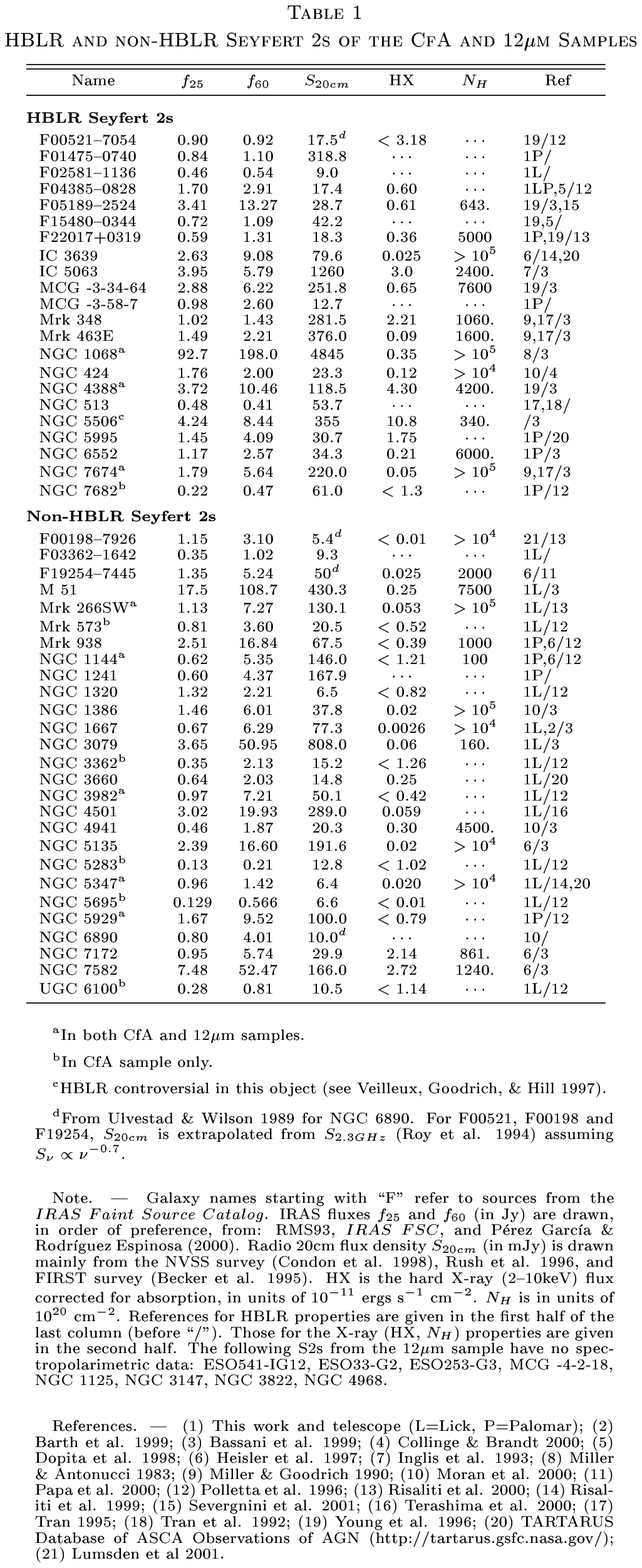}}
%\label{cfa12mtab}}

\noindent
of S2. We find that 15 objects in Table 2 of RMS93 are 
indeed LINERs, H II or SB galaxies misclassified as S2, 
and thus cannot be counted in the statistical analysis of Seyfert galaxies. In the course of the survey, 
most nevertheless have been observed. 
We will refer to these galaxies as the HLS sample (for H II, LINER, SB galaxies), and will present their data in 
paper II, but show them in plots in the current paper. 
Finally, one source, NGC 1194, is a S1 and will not be considered further.
The total number of S2s in the \tm~sample is thus 51, of which 43 have been observed either by this or other studies.  
Most of the remaining eight un-observed S2s are unreachable by telescopes employed in this survey.

In Table~1 we summarize the main results of our survey. 
The most relevant optical spectropolarimetric, X-ray, IR, and radio properties of the observed galaxies are presented. 
Besides confirming the spectropolarimetric properties of several galaxies, 
we discover one new HBLR S2 (NGC 7682) in the CfA sample, and four (F02581--1136, MCG -3-58-7, NGC 5995, NGC 6552)
in the \tm~sample. 

\section{Discussion} \label{disc}

\subsection{Two Populations of Seyfert 2s} \label{2pop}

Figure \ref{radirc} shows the two indicators of the relative power of the AGN: \frir~vs. \firr. 
The first measures the nonthermal radio emission normalized by the far-infrared emission (FIR, \fst), 
believed to arise mostly from circumnuclear SB and star formation (e.g., Alonso-Herrero et al. 2001; Ruiz et al. 2001). 
The second is the contribution of the mid-IR emission (\ftv), believed to come 
%FIG 1
\psfig{file=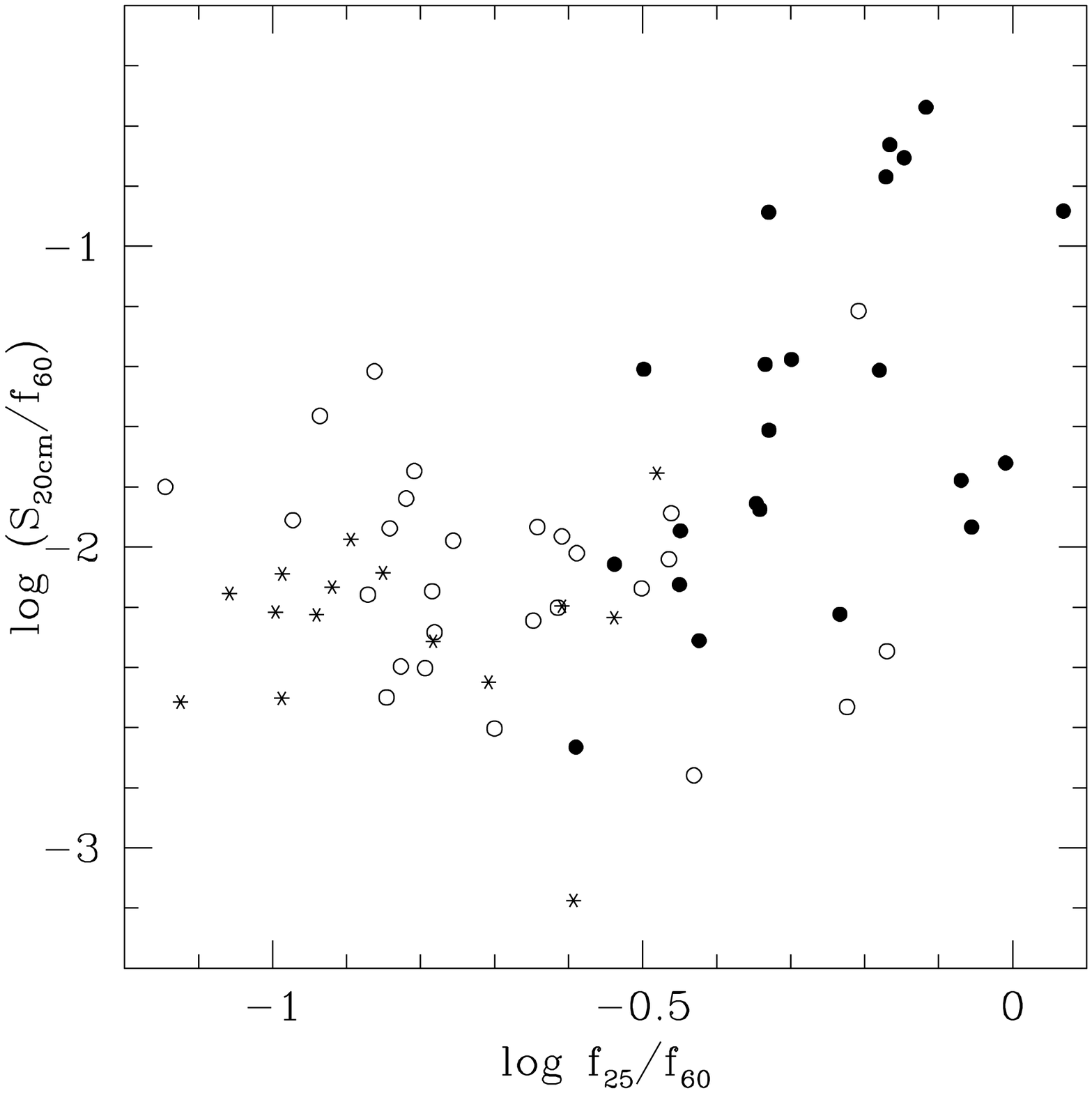,width=8.6cm}
%\psfig{file=fig1.eps,height=8.0cm}
%\hspace*{-10pt}
%\scalebox{0.45}{\includegraphics{fig1.eps}}
\figcaption{20cm radio flux density $S_{20cm}$, normalized by the FIR flux \fst,  
as a function of IR color \firr~for the CfA and \tm~samples. HBLR S2s are shown as solid dots, and 
non-HBLR S2s as open circles. Asterisks denote HLS galaxies, all of which have no HBLRs. 
For the HBLR S2s, as \frir~(AGN power) increases, \firr~(dust temperature) also increases. 
The same is not true for non-HBLR S2s.
The HBLR and non-HBLR S2s occupy distinctly different regions of the diagram, indeed showing markedly different trends, 
strongly suggesting that they are powered by intrinsically different AGN engines.
\label{radirc}}
\vskip 0.2cm

\noindent
mostly from surrounding dust (e.g., torus)
heated by the AGN \citep{rp97}, relative to the FIR emission. 
A separation is clearly detected, with {\it S2s identified as
having HBLR occupying a clearly distinct region of the plot compared to those without detection of HBLR.} 
The HBLR S2s tend to show both higher \frir~and \firr~ratios. This clearly marks them as truly energetic AGN, containing
a ``monster'' which is the hidden S1 nucleus. 
On the other hand, those S2s and LINERs/SB galaxies without HBLR present ``colder'', less energetic ratios: 
\firr~$\lesssim$ 0.25 and \frir~$\lesssim$ 0.01. 
The warmer infrared color for HBLR objects has been noted in numerous previous studies (e.g., Hutchings \& Neff 1991;
Heisler et al. 1997; Tran et al. 1999; Gu et al. 2001).
%\citet{hn91,t99,gu01}).
If all S2 are intrinsically similar, there is no {\it a priori} reason for them to lie in such distinct
regions of Figure \ref{radirc}.
\citet{t99} also advocate the use of another ``diagnostic diagram'' involving [O III] $\lambda$5007/H$\beta$~vs. 
\firr~to separate 
the HBLR from the non-HBLR S2s, with HBLR S2s showing generally higher ionization and warmer color. 
With the current combined CfA+\tm~sample, the median values of [O III] $\lambda$5007/H$\beta$~are 6.8$\pm$1.5 and 
9.9$\pm$1.3 for the non-HBLR and HBLR S2s, respectively. The corresponding values for \firr~are 0.18$\pm$0.10 and 
0.49$\pm$0.19, respectively, with virtually 0\% probability that they come from the same population. 
So the dichotomy is clear between these two populations. 

The question that arises is: is the lack of HBLRs in these galaxies 
due to the lack of an energetic AGN (and hence BLR), or to such impregnably high obscuration that we 
cannot even detect the signal from the buried AGN (i.e., $S_{20cm}$ and \ftv)? 
\citet{hlb97} suggested that the scattering may take place very close to the nucleus, 
in the inner ``throat'' of the torus, and non-HBLR S2s are perhaps those with the torus axes tipped 
at larger inclinations, resulting in higher obscuration 
of the nucleus and greater obstruction of the scattering region. 

%FIG 2
\hspace*{3pt}
\psfig{file=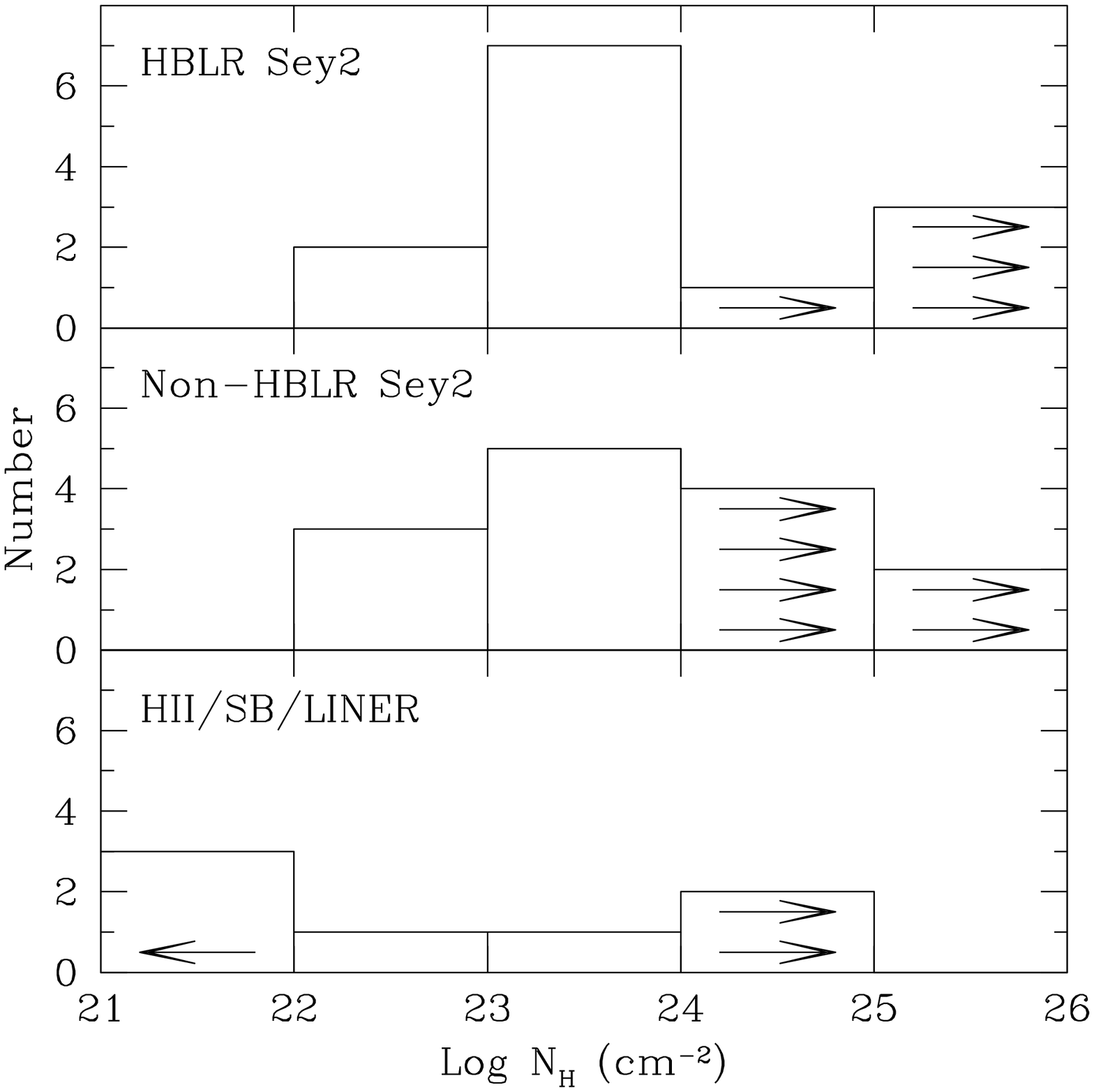,width=7.9cm}
\figcaption{The distribution of absorbing column density $N_H$ for S2s and HLS galaxies.
Arrows denote lower or upper limits. 
There is no significant difference in the mean $N_H$ for HBLR and non-HBLR S2s, indicating that non-HBLR S2s are not
any more obscured than HBLR S2s.
\label{nhdist}}
%\vskip 0.25cm

To answer the question posed above, we explore the level of obscuration in these galaxies in the X-ray regime.
Based on a smaller sample, \citet{al01} argued that the absorbing column density $N_H$ of HBLR and non-HBLR 
S2s do not show any significant difference. \citet{gu01}, on the other hand, suggested that HBLR S2s may have
on average smaller $N_H$ than non-HBLR S2s, using data available for a heterogeneous sample of HBLR S2s. 
For our larger and more complete sample we show in Figure \ref{nhdist} the distribution of $N_H$. 
Formally, the mean log($N_H$) values for the HBLR and non-HBLR S2s are 23.84$\pm$0.24 and 23.86$\pm$0.29, 
respectively -- virtually identical\footnote{The data have been treated with the ASURV package in IRAF, taking 
into account censored data.}! Arguably, the $N_H$ dataset is incomplete and subject to large uncertainties, 
and thus may not reflect the true nature of the obscuration in S2s. 

As another diagnostic of the obscuration, we consider the ratio of hard X-ray (2--10 keV) flux  
relative to the FIR flux, \hxir. Figure \ref{hxirc} displays the ratio \hxir~vs. \firr.
\hxir~is mainly a measure of the obscuration \citep{ris00,lev01}, but it also represents the strength of the 
buried AGN engine. Note that objects with higher obscuration show lower
\hxir~ratio, as expected, and objects with higher AGN power relative to the surrounding stellar radiation
lie toward the {\it right} of the diagram. The data are consistent with the model of \citet{ris00}, whose three 
trajectories corresponding to three representative values of $N_H$ are overplotted on Fig. \ref{hxirc}.
Pure S2s and HLS galaxies occupy the {\it bottom left} for all $N_H$.
For a similar range of $N_H$ (e.g., same color points), as the 
broad-line, energetic AGN component grows, the object moves to the {\it right}, and slightly up. 
%FIG 3
\hspace*{5pt}
\psfig{file=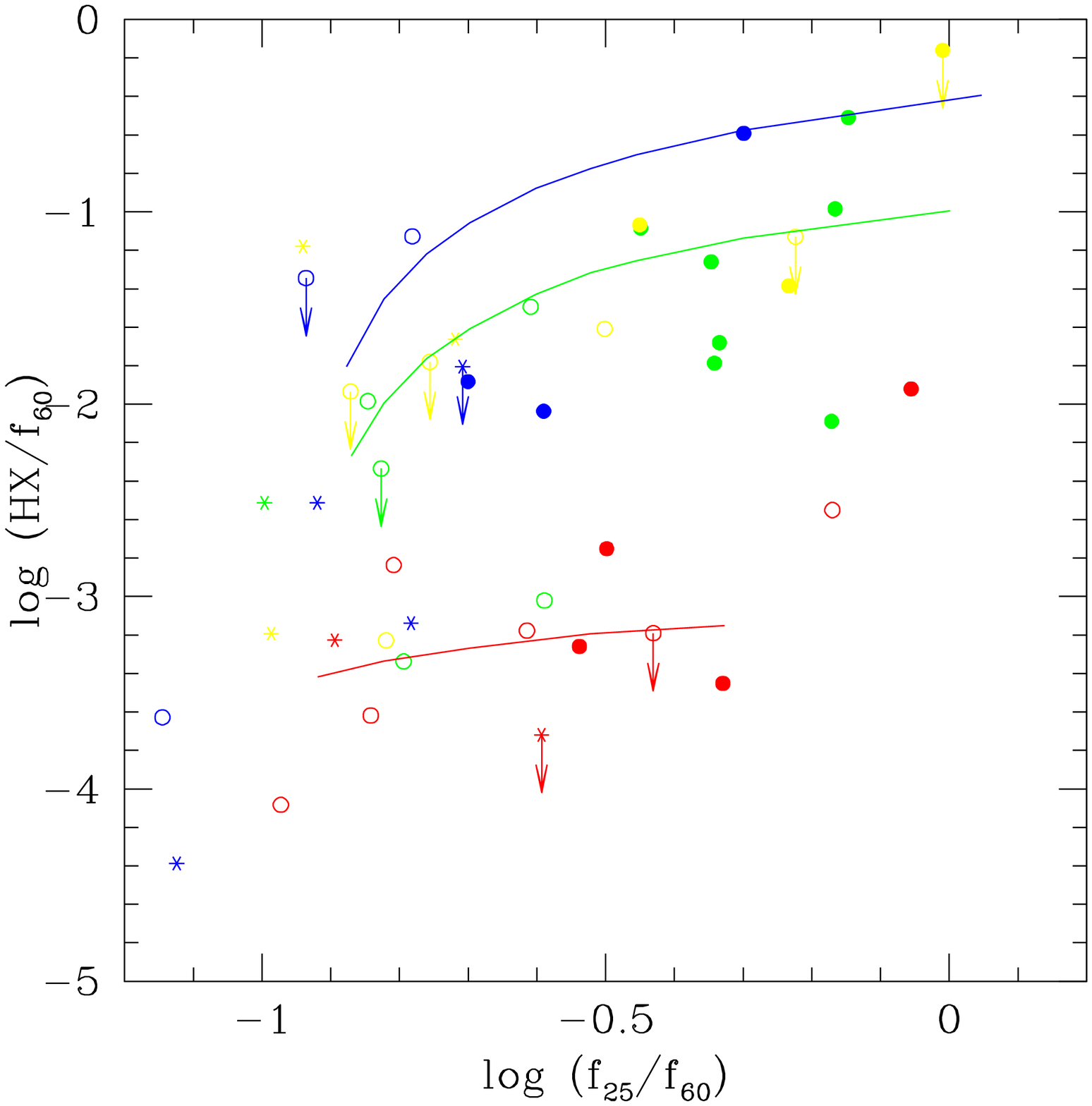,width=8.1cm}
\figcaption{The \hxir~ratio versus IR color \firr. Symbols are as in Figure \ref{radirc}. 
Color codes are as follows: red = $N_H > 10^{24}$ cm$^{-2}$ (Compton thick), 
green = $10^{23}~cm^{-2} < N_H < 10^{24}~cm^{-2}$, 
blue = $N_H < 10^{23}~cm^{-2}$, yellow = $N_H$ unknown. The red, green, and blue curves correspond to models
of Risaliti et al. (2000) for $N_H = 10^{24.6}$, $10^{23.6}$ and $10^{22}~cm^{-2}$, respectively. 
Were it not for the increased AGN strength in the
HBLR S2s, the HBLR and non-HBLR S2s would show the same range of \hxir~ratio, indicating similar levels of obscuration. 
Obscuration, therefore, does not play a great role in determining whether or not HBLR can be seen; 
the main factor seems to be AGN power. 
\label{hxirc}}
%\vskip 0.25cm

\noindent
Decreased obscuration then shifts it vertically up in the diagram. 
%(the reader is referred to the model of Risaliti et al. 2000 for details).
Close examination of Fig. \ref{hxirc} suggests that while the vertical offset is predominantly due to obscuration,
the slight trend upward to the top right can be attributed to increased AGN luminosity. 
Therefore, since the distributions of points with and without HBLR overlap much with each other over a broad 
range in \hxir, we conclude that {\it non-HBLR S2 are not any more obscured than HBLR S2}, consistent with the 
$N_H$ data, and in contrast with expectation from the \citet{hlb97} model. 
Figures \ref{nhdist} and \ref{hxirc} also show that the non-HBLR S2s and HLS galaxies do not show any evidence for 
higher level of obscuration than HBLR S2s, as would be expected if starburst activity, which tends to be higher in 
the former, provides an ``extra'' source of obscuration \citep{lev01} in addition to the torus. 

The Balmer decrement may also be used as a potential probe of the obscuration. For the current combined CfA+\tm~sample,
the mean \ha/\hb~is 7.16 $\pm$ 4.70 and 7.39 $\pm$ 5.51 for HBLR and non-HBLR S2s, respectively. KS test shows
that there is no statistical difference between these distributions, with a probability $p(null) = 0.36$ for the
null hypothesis that the two datasets are drawn from the same parent population. This confirms the similar 
X-ray result that there is no difference in obscuration between the two S2 types.
Our result disagrees with the suggestion of \citet{hlb97} and \citet{al01}, based on smaller, incomplete samples,
that non-HBLR S2s tend to display higher galactic extinction.

Thus the correlation in Figure \ref{radirc}, combined with the similarity in obscuration for both the HBLR and non-HBLR 
S2s lead us to our most important conclusion: {\it non-HBLR Seyfert 2s do not show HBLRs simply because
they most likely do not have any.} The UM model apparently is not applicable to {\it all} Seyfert galaxies. 
There appears to be two types of S2s , one containing an energetic AGN with BLR, and the 
other possessing a less energetic AGN with weak or absent BLR, whose energy output maybe dominated by other nuclear and 
circumnuclear processes such as starbursts. This is the ``pure S2'' model \citep{mor92,h95,dh99,gu01}.
 
As further support for the pure S2 model, very recent analysis of X-ray observations by \citet{pap01} shows that 
some S2s appear to show very little or no absorption, and yet display no broad lines, as would be expected if no 
obscuration is present.
There are also theoretical reasons to believe that intrinsically weak AGNs may 
lack BLRs. Nicastro (2000, see also Collin \& Hur\'e 2001) suggests that the BLR 
arises from the vertical disk wind from the accretion disk at a radius where 
it is unstable to radiation pressure. Below a certain critical accretion rate,
such wind, and hence BLR, cannot exist. Therefore, for sufficiently weak AGN, no BLRs are expected. 
The ``standard'' UM also cannot account 
for the apparent enhanced star forming activity \citep{mai95,gu98}, or higher frequency of companions \citep{dr98,dh99},
in S2 compared to S1 (see however, Schmitt et al. 2001). 
In the pure S2 model, the origin of these differences is clear: pure S2 galaxies are 
intrinsically different from S1; they reside in denser environments, which lead to a higher incidence of galaxy 
interactions, and hence star formation.

We emphasize that the lack of HBLR in the non-detected objects cannot be attributed to the overwhelming 
contribution of starlight in the host galaxies. Rather, it is the strength of the AGN engine that
seems to be dominant factor in determining the visibility of HBLR. Starlight level in HBLR S2s reaching $\sim$ 80-90\% 
is quite common \citep{t95,mor00}. Both HBLRs and non-HBLRs appear to have similar levels of starlight domination, 
and the non-detection of a HBLR seems to be unrelated to it \citep{t99,km98}. The fact that many starburst-dominated 
S2s do contain HBLRs (e.g., IC 3639, F05189-2524; Gonzalez Delgado et al. 2001; Cid Fernandes et al. 2001) also strongly 
suggests that the detectability of HBLR is not simply a function of the contribution of the starburst component 
relative to AGN power. For our sample, the mean 20cm radio powers in units of log solar luminosities for the 
HBLRs and non-HBLRs are 3.87 $\pm$ 0.60 and 3.10 $\pm$ 0.82, respectively. KS test shows that the 
two distributions are significantly different ($p[null]=0.015$), 
confirming a similar finding by \citet{mor92} for a limited number of objects. 
Very recently, Thean et al. (2001) also confirm that radio sources 
in the \tm~sample HBLR S2s do indeed contain more powerful radio sources.
So, not only are the relative radio to FIR fluxes different between the 
two types of S2s, the absolute radio luminosity also appears to be significantly different. 
   
\subsection{Differences between CfA and \tm~Samples} \label{diff}

The detection rate of HBLRs is significantly lower in the CfA sample (4/14 = 29\%) than the \tm~sample (21/43 = 49\%), 
although the CfA detection rate is similar to that reported by previous studies (e.g., Moran et al. 2000). The higher 
detection rate in the \tm~sample suggests that previous optically selected samples may have missed
many dust obscured AGNs. 
As argued by RMS93, the CfA sample appeared to be incomplete at both the 
high and low ends of the luminosity function\footnote{See also \citet{hu01} and \citet{the01} for further discussion 
of subtle selection effects in the CfA sample.}. According to their estimate, the incompleteness factor is about 50\%. 
Assuming that this translates directly to the same factor in the undercounting of HBLR S2s, it brings the detection 
rate of HBLRs in the CfA sample to $\sim$ 45\%, in much better agreement with the \tm~number. 
Thus, the $\approx$50\% detection rate of HBLR in S2s
may be more representative than the lower 30--35\% suggested by the CfA sample and other optically 
defined samples.

\acknowledgments
This research has been carried out through the years by HDT while at Caltech, Lick Observatory, LLNL, 
and the Johns Hopkins University. I would like to thank all my sponsors, M. H. Cohen, J. S. Miller, 
W. van Breugel and H. Ford, at these institutions for making it possible with their support. 
 
%\clearpage

\end{document}